\documentclass{article}
\usepackage{ijcai22}

\usepackage{times}
\usepackage{soul}
\usepackage{url}
\usepackage[hidelinks]{hyperref}
\usepackage[utf8]{inputenc}
\usepackage[small]{caption}
\usepackage{graphicx}
\usepackage{amsmath}
\usepackage{amsthm}
\usepackage{booktabs}
\usepackage{algorithm}
\usepackage{algorithmic}
\title{Macroeconomic forecasting and sovereign risk
assessment using deep learning techniques}

\author{
    Anastasios Petropoulos
    \affiliations
    Cyprus University  of Technology
    \emails
    apetropoulos@bankofgreece.gr
}

\author{
Anastasios Petropoulos$^1$
\and
Vassilis Siakoulis$^1$\and
Konstantinos P. Panousis$^{1}$\and
Loukas Papadoulas$^2$\And
Sotirios Chatzis$^1$
\affiliations
$^1$Cyprus University of Technology\\
$^2$Ethical AI Novelties\\
\emails
\{apetropoulos, vsiakoulis\}@bankofgreece.gr,
k.panousis@cut.ac.cy, \\
luke@ethicalaicy.com, sotirios.chatzis@cut.ac.cy
}

\begin{document}

\maketitle

\begin{abstract}
    In this study, we propose a novel approach of nowcasting and forecasting the
    macroeconomic status of a country using deep learning techniques. We focus particularly on
    the US economy but the methodology can be applied also to other economies. Specifically
    US economy has suffered a severe recession from 2008 to 2010 which practically breaks out
    conventional econometrics model attempts. Deep learning has the advantage that it models
    all macro variables simultaneously taking into account all interdependencies among them
    and detecting non-linear patterns which cannot be easily addressed under a univariate
    modelling framework. Our empirical results indicate that the deep learning methods have a
    superior out-of-sample performance when compared to traditional econometric techniques
    such as Bayesian Model Averaging (BMA). Therefore our results provide a concise view of a
    more robust method for assessing sovereign risk which is a crucial component in investment
    and monetary decisions.
\end{abstract}

\section{Motivation}
    Deep Learning models have a short history, however, they have found application in a variety of scientific fields. In particular, Deep Learning algorithms have dramatically improved the capabilities of performing pattern recognition (like speech recognition, image recognition) and forecasting, so that they offer state of the art performance in various scientific fields like biology, engineering etc. Their structure offers the ability to adjust in streaming sequences using continuous learning algorithms, and recognize new and evolving patterns in time series data. In addition, deep learning is proven to effectively deal with high dimensional data. Their capacity to learn and adapt to new data can lead to a better predictive performance in financial time series modelling problems where non-linear relationships and observational noise often exist.
    
    Furthermore, this new generation of statistical algorithms offers the necessary flexibility in modelling multivariate time series, as its structure includes a cascade of many layers with non-linear processing agents. Deep learning networks base their functionality on the interaction of layers that simulate the abstraction and composition functionalities of the human brain. Therefore, via capturing the full spectrum of information contained in financial datasets, they are capable of exploring in depth the inherent complexity of the underlying dynamics in big and high dimensional time series data.
    
    In this study we explore a novel approach to now-casting and forecasting the macroeconomic status of a country using deep learning techniques. We focus particularly on the US economy but the methodology can be applied also to other economies. Specifically US economy has suffered a severe recession from 2008 to 2010 which practically breaks out single time series model attempts. Our approach can simultaneously simulate a country’s key financial variables in a holistic way, under a dynamic balance sheet assumption, and by utilizing deep learning algorithms. Experimental results give strong evidence that deep learning applied in financial datasets creates a state of the art paradigm, which is capable to simulate real word scenarios in a holistic and more efficient way. Deep learning has the advantage that it models all macro variables simultaneously taking into account all interdependencies among them and detecting non-linear patterns which cannot be easily addresses under univariate modelling framework.
    
    The main contribution of this study is that it proposes a holistic framework for macroeconomic forecasts. Our research analysis lies at the intersection of computational finance and statistical machine learning, leveraging the unique properties and capabilities of deep learning networks to increase the prediction efficacy and minimizing the modelling error. Under the proposed approach, forecasting of macroeconomic situation can be heavily supported by artificial intelligence algorithms simulating better the propagation channels across different parts economy i.e. banking system, consumer confidence and state interaction. In a nutshell, the proposed innovation lies in the use of advanced deep learning techniques for the simultaneous projection of Macroeconomic variables, while benchmarking vs with traditional methods of econometrics frequently employed in financial practice (i.e. Bayesian Model averaging). 
    
    The remainder of this study is structured as follows. In section 2, we focus on the related literature review. Section 3, describes the data collection and processing. In section 4, we provide details regarding the estimation process of the various alternative models developed.  In section 5, we compare the employed methodologies. Finally, in  concluding section 6, we summarize our findings and identify any potential weaknesses or limitations, while we also discuss areas for future research extensions.

\section{Related Work}

    Macroeconomic Forecasting is a very useful tool effectively utilized in banking, finance, business, and other areas. Academic studies have established a close  link between sovereign risk and the expected GDP growth of a country. Vice versa a country’s economic growth exhibits a significant response to sovereign risk changes driven by the interest rate and capital-flow channels\cite{chen2016relation}. Furthermore, the borrowing costs of the economy are statistically and economically affected by an increase in sovereign risk as measured through firm’s credit spreads, based on a recent analysis of credit default swap data in the Eurozone\cite{bedendo2015sovereign}. 
	
    Macroeconomic satellite models are also the cornerstone of bank stress testing methodologies as they are the pipeline through which macroeconomic scenarios are converted to tangible risk factors. In this framework macro models are used either for scenario projections under specific assumptions or dynamic forecasting. Usually banks develop a group of satellite models each for every one macro variable based on which projections and mapping to risk factors are performed. Multi models usually ignore the interrelations among variables except for the cases where a joint distribution (copula) is employed for correlation modelling. Underestimating the correlation impact may adversely affect the validity of stress test results as second order effects deployed during a crisis period will be ignored and the estimated impact will deviate from realized losses.
	
	With respect to sovereign risk assessment, the majority of the academic literature employs conventional econometric and statistical techniques to tackle the problem of government debt credit risk assessment. These methods range from simple regression model to classification-regression trees\cite{manasse2009rules}.
	
	The recent developments of Machine and Deep Learning methodologies and the availability of large scale macroeconomic data have led a number of researchers to depart from the field of traditional econometric techniques and employ novel techniques both in forecasting and now-casting macroeconomic variables.  Bontempi et al.\cite{Bontempi2013} make an overview of machine learning techniques in time series forecasting by focusing on three aspects: the formalization of one-step forecasting problems as supervised learning tasks, the discussion of local learning techniques as an effective tool for dealing with temporal data and the role of the forecasting strategy when we move from one-step to multiple-step forecasting. Atiya et al.\cite{ahmed2010empirical}, perform a large scale comparison study for the major machine learning models for time series forecasting by applying the models to the monthly M3 time series competition data. Katris .\cite{katris2020prediction} benchmarks machine learning techniques vs traditional econometric models in the prediction of Unemployment Rates.
	
	Liao\cite{liao2017machine} used Artificial Neural Network in time-series forecasting, by combining First order Markov Switching Model and K-means algorithms and found that machine learning has outperformed the benchmark of time-series inflation rate forecasting.  Medeiros et al.\cite{medeiros2021forecasting} showed that ML models with a large number of covariates are systematically more accurate than the benchmarks in US inflation since they capture better potential nonlinearities between past key macroeconomic variables and inflation. Smalter\cite{hall2018machine} points that supplied with diverse and complex data, a machine learning model can outperform simpler time-series models, with better performance at shorter horizons. In particular, his results show that a machine learning model can identify turning points in the unemployment rate earlier than competing methods. 
	
	Kaushik et al.\cite{kaushik2020forecasting} come up with a multivariate time series approach to forecast the exchange rate. His results show that Support Vector Machines and Recurrent Neural Networks outperform widely used traditional method of econometric forecasting such as Vector Autoregressive Models. In a similar vein, Chen et al. \cite{chen2004regression} employ a two stage model in exchange rate forecasting where in the first stage, a time series model generates estimates of the exchange rates whereas in the second stage, General Regression Neural Network is used to correct the errors of the estimates. Both empirical and trading simulation experiments suggest that the proposed hybrid approach not only produces better exchange rate forecasts but also results in higher investment returns than the single-stage models.

\section{Data Collection and Model Structures}
	
	In order to calibrate a Deep Neural Network to a series of macroeconomic series we have collected monthly series from macroeconomic and financial indicators for the US economy from January 1973 to December 2018. The described dataset comprised of 540 observations and was split into two parts: An in-sample train dataset, comprising data pertaining to the $65\%$ (1973-2005) and an out-of-sample dataset that comprises the rest $35\%$ of the observations (2006-2018). The latter sample was used for evaluation of the performance of all the statistical techniques implemented. In the split of the sample the 2008-2010 crisis period was deliberately left out in order to challenge our model capacity in foreseeing an exceptional structural break in the macroeconomic status. Thus, the developed system is expected to exhibit more stability through the cycle behaviour.
	The macroeconomic series incorporated into the network depict all relevant aspects of the economic status of a country. Namely we include in the network the following 9 macroeconomic time series:
	
	\begin{itemize}
		\item GDP: Gross Domestic Product yearly growth
		\item DEBT:  Government Debt as \% of GDP
		\item  RRE: Real Estate prices yearly growth
		\item UNR: Yearly change in Unemployment Rate
		\item INFLAT: Price inflation i.e. Consumer Price Index yearly growth
		\item YIELD10Y: 10 year Government bond yield changes
		\item GOVEXP: Government Expenses as \% of GDP
		\item EXPORT: Exports yearly growth
		\item STOCKS: Annual return of S\&P 500
	\end{itemize}
	
	Our aim is to effectively forecast the evolution of the abovementioned time series both under a univariate framework by employing a Bayesian Moving Average model and under a multi-variate framework employing Deep Neural Networks taking into account simultaneous interdependencies. In the next step we benchmark the forecasting performance of the univariate vs the multivariate framework both under a static and dynamic view.
	We employ an explanatory set of variables for each of the 9 macro series 1 to 2 yearly lags of both the dependent and the rest of the other macro series based on the assumption that macro and financial variables ratios carry all the information necessary to describe and predict the macroeconomic state of a country. We also include as independent variables a set of variables, on current and lagged values, which account for the effect of banking system in the real economy, namely
	
	\begin{itemize}
		\item DEP: Bank Deposits yearly growth
		\item LOAN: Commercial Loans yearly growth
		\item INTLOAN: Interbank loan yearly growth
	\end{itemize}

	This is a crucial supplement as the interlinkage between banking sector and economic activity is both empirically and theoretically justified. Confidence of the public and the market in the banking sector will lead to liquidity provision through the disposition of customer deposits and interbank funding which in turn will be to commercial loans which will fuel economic activity. On the other hand, low banking confidence leads to deposit outflows and shrinkage of the interbank market, rendering problematic the financing of the economy from the banking system. This is the main reason why a crisis in the banking sector could lead to an economic crisis which in turn could send feedback effects on the banking system initiating a new second order crisis.  
	In all, the individual models estimated through Bayesian Model Averaging are shown below in eq (1).
	
	\begin{align}
	\begin{split}
		\text{DEPVAR}_t &= \text{GDP}_{t-1,2} + \text{DEBT}_{t-1,2} + \text{RRE}_{t-1,2} + \text{UNR}_{t-1,2}\\
		& + \text{INFLAT}_{t-1,2} + \text{YIELD10Y}_{t-1,2} + \text{GOVEXP}_{t-1,2}\\
		& + \text{EXPORT}_{t-1,2} + \text{STOCKS}_{t-1,2} + \text{DEP}_{t-0,1,2}\\
		& + \text{LOAN}_{t-0,1,2} + \text{INTLOAN}_{t-0, 1,2} + \epsilon
	\end{split}
	\end{align}
	where $\text{DEPVAR}_t$ comprises each one of the following variables ($\text{GDP}_t$, $\text{DEBT}_t$, $\text{RRE}_t$, $\text{UNR}_t$, $\text{INFLAT}_t$, $\text{YIELD10Y}_t$, $\text{GOVEXP}_t$, $\text{EXPORT}_t$, $\text{STOCKS}_t$) leading to 9 separate models, one for each macroeconomic factor.
	
	Modeling separately the macroeconomic variables has the drawback that it does not take into account the simultaneous interdependencies among variables. For example it could the case that GDP is not affected solely by the lagged value of RRE but also by the $\text{RRE}_t$. Traditional econometrics handle this case through the use of SUR (Seemingly Unrelated Regression) models or under a VAR framework. In our proposed approach for dynamic macro forecasting we introduce a Deep neural networks architecture as an innovative way to take into account contemporaneous dependencies among variables. 
	
	In all, we identify the main channels of risk propagation in a recurrent form to account of all the existing evidence of feedback effects in a macroeconomic system by putting all the components together in a multivariate structure. On the other hand, the use of classical econometric techniques offers limited capabilities for simulating complex systems. Our approach accounts for temporal patterns in the economy providing a dynamic modelling approach. This is achieved through the multivariate training of deep neural networks which takes into account the dynamic nature of the economy. The approach proposed is composed of multivariate input and output layers able to capture the cross correlation between macroeconomic variables. Training is performed as one big complex network minimizing estimation errors and double counting effects among various financial variables.
	
	To account for non-linear relationships that materialize under different macroeconomic conditions machine learning techniques like deep learning can provide more efficient estimations. Based on academic literature Deep Neural networks are capable of simulating real life phenomena where relationships are complex, so our proposed framework by using multilayer deep networks envisages capturing the dynamics inherent in the economy. In addition the architecture of aims to capture the amplification channels leading to structural breaks. 
	
	In the general structure of our model lagged macroeconomic indicators along with current values for Loan Growth, Deposit Growth and Interbank Loan growth will be inserted in the input layers and their now-casted values will be produced in the output layer.

\section{Model Development}
	
   \subsection{Single Variable Forecasting – Bayesian Model Averaging (BMA)}
	
	We employ the Bayesian Model Averaging (BMA) methodology for estimating single equation satellite models which are used for univariate predictions for the macroeconomic variables. The BMA method accounts better than linear regression for the uncertainty surrounding the main determinants of risk dynamics. Using BMA, a pool of equations is generated using a subgroup of determinants randomly selected. In the next step, a weight is assigned to each model that reflects their relative forecasting performance. Aggregating all equations using the corresponding weights produces a posterior model probability, provided that the number of equations estimated in the first step is large enough to capture all possible combinations of a predetermined number of independent variables. Thus Bayesian model averaging addresses model uncertainty and misspecification better than a simple linear regression problem.
	
	To further illustrate BMA, suppose a linear model structure with $Y_t$ being the dependent variable, $X_t$ the explanatory variables, $\alpha$ constant, $\beta$ the coefficients, and $\epsilon_t$ a normal error term with variance $\sigma$.
	
	\begin{align}
		Y_t = \alpha_\gamma + \beta_\gamma X_{\gamma, t} + \epsilon_t,
		 \qquad \epsilon_t \sim \mathcal{N}(0,\sigma^2 I )
	\end{align} 
	
	In an ordinary linear regression problem, the existence of many potential explanatory variables in a matrix $X_t$ renders their correct combination a quite burdensome task. Including all the variables does not provide a feasible solution as it can lead to overfitting and multicollinearity especially when there is a limited number of observations. BMA tackles the problem by estimating models for all possible combinations of $\{X\}$ and constructing a weighted average over all of them.
	
	Under the assumption that $X$ contains $K$ potential explanatory variables, BMA estimates $2^K$ combinations and thus $2^K$ models. Applying Bayes’ theorem, model averaging is based on the posterior model probabilities.
	\begin{align}
	\begin{split}
	p(M_\gamma \cup Y, X) &= \frac{p(Y \cup M_\gamma, X) p(M_\gamma)}{p(Y \cup X)} \\
	&= \frac{p(Y \cup M_\gamma, X) p(M_\gamma)}{\sum_{s=1}^{2^K} p(Y \cup M_s, X) p(M_s)}
	\end{split}
	\label{eqn:bms_posterior}
	\end{align}
	In Equation \eqref{eqn:bms_posterior}, $p(Y,X)$ denotes the integrated likelihood which is constant over all models and is thus simply a multiplicative term. Therefore, the posterior model probability (PMP) is proportional to the integrated likelihood $p(Y \cup M, X)$ which reflects the probability of the data given the model $M$. Thus, the corresponding weight assigned to each model is measured by using $p(M_\gamma \cup Y, X)$ in Eq. \eqref{eqn:bms_posterior}.
	In equation \eqref{eqn:bms_posterior}, $p(M)$ denotes the prior belief of how probable model $M$ is before analyzing the data. Furthermore, to estimate $p(Y,X)$ integration is performed across all models in the model space and to estimate the probability $p(Y \cup M, X)$  integration is performed given model $M$ across all parameter space. By performing renormalization of the product in equation \eqref{eqn:bms_posterior}, PMPs can be inferred and subsequently the model's weighted posterior distribution for estimator $\beta$ is given by
	\begin{align}
	p(\beta \cup Y, X) = \sum_{\gamma=1}^{2^K} p(\beta \cup M_\gamma, Y, X) p(M_\gamma \cup X, Y)
	\end{align}
	The priors, posteriors and the marginal likelihood employed in the estimation are described analytically in Appendix \ref{appendixa}. In Bayesian Model Averaging estimation we employ unit information prior (UIP), which sets g=N commonly for all models. We use also a birth/death MCMC algorithm (20000 draws) due to the large number of covariates included since using the entire model space would lead to a large number of iterations. We fix the number of burn-in draws for the MCMC sampler to 10000. Finally, the models prior employed is the ``random theta'' prior by Ley and Steel, who suggest a binomial-beta hyperprior on the a priori inclusion probability. This has the advantage that is less tight around prior expected model size (i.e. the average number of included regressors) so it reflects prior uncertainty about model size more efficiently. In order to develop all the satellite models for this approach we employ the utilities of BMS R package. 
	
  \subsection{Multiple Variable Forecasting - Deep Neural Networks}

	We propose the use of multilayer deep neural networks in order to simultaneously forecast the basic macroeconomic and financial variables, capturing the dynamics inherent in the economy by taking into account the contemporaneous interdependencies among them. Deep learning applications in the domain of finance is rather limited but it has been an active field of research in recent years, as it has achieved significant breakthroughs in the fields of computer vision and language understanding. Specifically, our paper constitutes one of the first works presented in the literature that considers application of deep learning to address the challenging task of macroeconomic prediction. 

Deep Neural Networks are built on the basis of nonlinear activation functions typical choices of which are the logistic sigmoid, hyperbolic tangent, and rectified linear unit (ReLU). The first two (logistic sigmoid and hyperbolic tangent) activation functions are closely related as they belong both to the sigmoid family. The sigmoid family of activation functions has the disadvantage of saturating the network with large positive or negative values. To alleviate this problem, practitioners have derived linear activation functions, like the popular ReLU, which are now the standard choice in deep learning research. 
	
	The activation layers increase the ability and flexibility of a DNN to capture non-linear relationships in the training dataset but on the other hand, the huge number of trainable parameters could lead to overfitting. Therefore the use of simple, effective, and efficient regularization techniques is necessary to avoid poor out of sample performance. Dropout is the most popular regularization technique for DNNs. In essence, it consists in randomly dropping different units of the network on each iteration of the training algorithm, so in this way, only the parameters related to a subset of the network units are trained during each iteration. This strategy reduces the associated network overfitting tendency ensuring that all network parameters are effectively trained. Inspired by these merits, we employ Dropout DNNs with ReLU activations to train and deploy feed forward deep neural networks. We postulated deep networks that are up to five hidden layers deep and comprise various numbers of neurons. 
	
	A supplementary way to improve the fitting efficacy of the trained network is the incorporation prior information in a Bayesian framework in a similar way to BMA linear models. Conventional network architectures compute point estimates of the unknown values without taking into consideration any prior information and without any uncertainty estimation of the produced values. The Bayesian treatment of a particular model has been shown to increase its capacity and potential, while oﬀering a natural way to assess the uncertainty of the resulting estimates. To this end, we augment the conventional model architectures of the previous sections by also relying on the Bayesian framework. 
	
	Specifically, we impose a prior Normal distribution over network weights, seeking to infer their posterior distribution given the data. Since the marginal likelihood is intractable for the considered architectures, from the existing Bayesian methods, we rely on approximate inference and specifically on Variational Inference. Since the true posterior of the model cannot be computed in Variational Inference, we introduce an auxiliary variational posterior distribution of a family of distributions and then try and find the optimal member of the considered family to match the true posterior. The matching is achieved through the minimization of the Kullback-Leibler (KL) divergence between the true and the introduced variational posterior{\protect\footnote{The KL divergence is a metric of similarity between two distributions and is a non-negative value; KL is zero, if and only if, the two considered distributions match exactly.}}. Minimizing the KL divergence is equal to the maximization of the Evidence Lower Bound (ELBO), a well-known bound on the marginal likelihood derived using Jensen’s inequality. Thus, for training the following architectures, we resort to ELBO maximization.
	
	Non-linear activation functions such as ReLUs are a mathematically convenient tool for training deep networks but nevertheless, they do not come with strong biological plausibility. Current research has shown that in real life neurons with similar functionality and structure tend to group and compete with each other for their output. To this end, researchers have devoted signiﬁcant effort to explore this type of competition between neurons and apply it in existing models. The resulting procedure is referred to as Local Winner-Takes-All (LWTA) and has been shown to provide competitive, or even better, results in benchmarks architectures in different domains of Deep Learning applications. Thus, apart from the conventional ReLU activations of the previous section, we additionally explore the potency of the LWTA in our work. The linear units after the affine transformation in each layer are grouped together and compete from their outputs. This competition is performed in a probabilistic way, by employing a softmax nonlinearity, obtaining the probability of activation of each unit in each block.

\section{Model Validation}
	
	In order to assess the robustness of our approach we perform a thorough validation procedure. The dataset comprised of 540 observations was split into two parts: An in-sample train dataset, comprising data pertaining to the 65\% (1973-2005) and an out-of-sample dataset that comprises the rest 35\% of the observations (2006-2018). The latter sample was used for evaluation of the performance of all the statistical techniques implemented. In the split of the sample, the 2008-2010 crisis period was deliberately left out in order to challenge our model capacity in foreseeing an exceptional structural break in the macroeconomic status. Models compared in this benchmark exercise are Bayesian Model Average (BMA), Deep Learning Model with ReLU activation function and Drop out terms trained using the MXNET algorithm (MXNET), Bayesian Deep Learning model using ReLU activation function (Bayesian ReLU) and  Bayesian Deep Learning model using Local Winner-Takes-All activation function (Bayesian LWTA).
	
	Note that, in order to train the deep learning algorithms in the current study, the in-sample dataset is further split randomly in train and validation set. The validation dataset is used to find the best set of hyper parameters of the models and select the best candidate model for performing the out–of-sample evaluation.  
	
	The validation process is performed following two different strategies, namely static and rolling forecast. In the static case, all models are estimated only one time in the train sample (1973-2005) and the whole test sample is used to produce the out of time (2006-2018) predictions in a one-off way. In the rolling forecast framework the models are re-estimated in each step of the out of time sample (2006-2018) and the prediction is produced for one month ahead each in each step. For example in order to produce the January 2006 forecast the sample 1973 till December 2005 is used, whereas in the next step the train sample is enlarged to 1973 till January 2006 and the February 2006 forecast is generated. 
	
	From Table \ref{table:static} where the errors for each forecasted macro variable are shown we deduce that Deep Learning algorithms clearly outperform the benchmark BMA approach, especially in cases of variables such as the government bond yield (YIELD10Y), the real estate price growth (RRE) and the stock market growth (STOCKS) which showed a massive structural outbreak during the Subprime crisis (2008-2010) in the US. The Bayesian Deep Neural Network with Local Winner Takes All activation function outperforms not only the BMA model but also more conventional Deep Learning algorithms showing the benefit of applying more biologically plausible activation functions. 

	\begin{table}[!h]
		\centering
		\caption{Static Forecast Error Metrics for each variable: MSE stands for Mean Square Error and MAE stands for Mean Absolute Error. }
		\label{table:static}
		\renewcommand{\arraystretch}{1.1}
		
		\resizebox{\linewidth}{!}{
			
		\begin{tabular}{|c|c|c|c|c|c|c|}
			\hline
			Method & \multicolumn{2}{c|}{\textbf{YIELD10Y}} & \multicolumn{2}{c|}{\textbf{UNR}} & \multicolumn{2}{c|}{\textbf{RRE}}\\\hline
			 & MSE & MAE & MSE & MAE & MSE & MAE \\\hline
			Satellite Modeling (BMS) & 13.31 & 2.54 & 1.40 & 0.90 & 79.14 & 7.09\\\hline
			Deep Learning (MXNET) & 1.14 & 0.81 & 1.95 & 0.97 & 40.42 & 5.21 \\\hline
			Deep Learning (Bayesian ReLU) & 1.76 & 0.93 & 0.95 & 0.73 & 36.78 & 4.83 \\\hline
			Deep Learning (Bayesian LWTA) & \textbf{1.08} & \textbf{0.80} & \textbf{0.87} & \textbf{0.69} & \textbf{28.56} & \textbf{4.33} \\\hline
			& \multicolumn{2}{c|}{\textbf{INFLAT}} & \multicolumn{2}{c|}{\textbf{STOCKS}} & \multicolumn{2}{c|}{\textbf{GDP}}\\\hline
			 & MSE & MAE & MSE & MAE & MSE & MAE \\\hline
			 Satellite Modelling(BMA) & 6.73 & \textbf{1.90} & 472.83 & 17.32 & 8.80 & 2.42\\
			 Deep Learning (MXNET) & 12.61 & 2.95 & \textbf{158.55} & \textbf{8.69} & 4.55 & 1.74\\
			 Deep Learning (Bayesian ReLU) & 10.14 & 2.46 & 172.06 & 9.26 & \textbf{ 4.30} & \textbf{1.68}\\
			 Deep Learning (Bayesian LWTA) & \textbf{5.51} & 1.92 & 171.01 & 9.86 & 6.58 & 2.00\\\hline
			 & \multicolumn{2}{c|}{\textbf{EXPORT}} & \multicolumn{2}{c|}{\textbf{DEBT}} & \multicolumn{2}{c|}{\textbf{GOVEXP}}\\\hline
			 & MSE & MAE & MSE & MAE & MSE & MAE \\\hline
			 Satellite Modelling(BMA) & 159.57 & 9.73 & 13.30 & 2.63 & 14.2 & 1.47\\
			 Deep Learning (MXNET) & 103.81 & \textbf{7.88} & 16.53 & 2.73 & 2.28 & 1.18\\
			 Deep Learning (Bayesian ReLU) & 100.36 & 8.10 & \textbf{12.10} & \textbf{2.41} & \textbf{1.31} & \textbf{0.84}\\
			 Deep Learning (Bayesian LWTA) & \textbf{93.70} & 8.01 & 18.57 & 2.93 & 2.32 & 1.16\\\hline
			 
		\end{tabular}}
	
	\end{table}
 
\begin{figure}[!h]
    \centering
    \includegraphics[width=\linewidth]{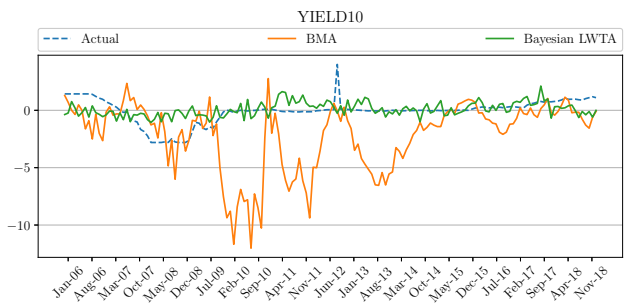}
    \caption{10 year Government bond yield change- Forecast Comparison. Actual value vs Bayesian Model Average (BMA) vs Bayesian Neural Network with Local Winner-Takes-All activation function (Bayesian LWTA).}
    \label{fig:yield10}
\end{figure}

	In order to compare the prediction of the two models with actual value of each variable, the diagrams shown in the Appendix have been created. In particular, from the graph in Figure \ref{fig:yield10} , which shows the scaling of the actual value of YIELD10Y from August 2016 to November 2018, it can be clearly seen that the Bayesian LWTA follows the trend of the actual value much better than the BMA, which deviates significantly, given that it does not take into account the contemporaneous interdependencies among variables in the long run. Of course, there are cases such as Government Expense and Debt to GDP where Deep Learning model also overshoots but those are variables heavily dependent on political decisions which may diverge from the pattern observed especially in crisis periods. One should also take into account the challenging nature of the testing sample which includes a structural break (Subprime crisis) that has not been observed in the train sample, during which important fiscal decisions were taken from the US government in order to mitigate the crisis repercussions.

\newpage

	Focusing on the rolling forecast experiment in Table \ref{table:rolling}, we notice, as expected, that errors have reduced significantly vs the static case as more current information space is available in each step of the forecasting process. If the static experiment represents a situation of forecasting a macro variable full path on the basis of a model and a scenario provided, the rolling forecast experiment simulates a situation of a practitioner that re-trains a model on a monthly basis  so as to have a hint on what will happen in the next month. From a benchmarking perspective even in the rolling forecast case Bayesian Deep Learning algorithms clearly outperform the BMA model in all variables as they better account for contemporaneous relationships across variables. 
	
	By examining Figs. \ref{fig:yeild10rolling} to \ref{fig:GDPprolling} in Appendix B we notice that both BMA and the best performing Deep Learning algorithm (Bayesian ReLU) follow closer the trend in dependent variables evolution but the former in many cases overshoots rendering the Deep Learning alternative more robust for macroeconomic projections under a crisis situation.
	
	\begin{table}[!h]
		\centering
		\caption{Rolling Forecast Error Metrics for each variable: MSE stands for Mean Square Error and MAE stands for Mean Absolute Error.  }
		\label{table:rolling}
		\renewcommand{\arraystretch}{1.1}
		
		\resizebox{\linewidth}{!}{
			
			\begin{tabular}{|c|c|c|c|c|c|c|}
				\hline
				Method & \multicolumn{2}{c|}{\textbf{YIELD10Y}} & \multicolumn{2}{c|}{\textbf{UNR}} & \multicolumn{2}{c|}{\textbf{RRE}}\\\hline
				& MSE & MAE & MSE & MAE & MSE & MAE \\\hline
				Satellite Modeling (BMS) & 3.34 & 1.29 & 0.4 & 0.48 & 18.02 & 3.32\\\hline
				Deep Learning (MXNET) & 1.34 & 0.87 & 0.74 & 0.66 & 29.40 & 4.34 \\\hline
				Deep Learning (Bayesian ReLU) & \textbf{0.50} & \textbf{0.49} & \textbf{0.09} & \textbf{0.23} & \textbf{15.50} & \textbf{3.08} \\\hline
				Deep Learning (Bayesian LWTA) & 1.44 & 0.96 & 0.26 & 0.38 & 22.16 & 3.77 \\\hline
				& \multicolumn{2}{c|}{\textbf{INFLAT}} & \multicolumn{2}{c|}{\textbf{STOCKS}} & \multicolumn{2}{c|}{\textbf{GDP}}\\\hline
				& MSE & MAE & MSE & MAE & MSE & MAE \\\hline
				Satellite Modelling(BMA) & 3.07 & 1.30 & 203.02 & 11.43 & 2.59 & 1.26\\
				Deep Learning (MXNET) & 5.40 & 1.82 & \textbf{144.25} & 9.29 & 2.59 & 1.28\\
				Deep Learning (Bayesian ReLU) & \textbf{2.96} & \textbf{1.07} & 151.93 & \textbf{7.99} & \textbf{0.48} & \textbf{0.51}\\
				Deep Learning (Bayesian LWTA) & 4.62 & 1.49 & 157.06 & 8.49 & 3.25 & 1.38\\\hline
				& \multicolumn{2}{c|}{\textbf{EXPORT}} & \multicolumn{2}{c|}{\textbf{DEBT}} & \multicolumn{2}{c|}{\textbf{GOVEXP}}\\\hline
				& MSE & MAE & MSE & MAE & MSE & MAE \\\hline
				Satellite Modelling(BMA) & \textbf{72.56} & \textbf{6.96} & 6.98 & 2.05 & 14.40 & 1.12\\
				Deep Learning (MXNET) & 92.16 & 7.43 & 14.95 & 2.72 & 1.10& 0.82\\
				Deep Learning (Bayesian ReLU) & 87.93 & 7.26 & \textbf{2.26} & \textbf{1.13} & \textbf{0.40} & \textbf{0.50}\\
				Deep Learning (Bayesian LWTA) & 102.19 & 8.18 & 2.99 & 1.42 & 1.14 & 0.90
				\\\hline
				
		\end{tabular}}
		
	\end{table}

\section{Conclusions}
	
	In this study we explore a novel approach of now-casting and forecasting the macroeconomic status of a country using deep learning techniques focusing particularly on the US economy but the methodology can be applied also to other countries. Deep Learning algorithms have a short history in economics but this new generation of statistical algorithms offer the necessary flexibility in modelling multivariate time series, as its structure includes a cascade of many layers with non- linear processing agents.
	
	In particular, we identify the main channels of risk propagation in a recurrent form to account of all the existing evidence of feedback effects in a macroeconomic system by putting all the components together in a multivariate structure. Our approach takes into account the dynamic nature of the economy, through the multivariate training of deep neural networks, that employ multivariate input and output layers which are able to capture the cross correlation between macroeconomic variables. Training is performed as one big complex network minimizing estimation errors and double counting effects among various financial variables.
	Benchmarking a series of Deep Learning algorithms versus Bayesian Model regressions on a test sample that includes a financial turbulent period in the US (2008 – 2012) we find that Deep Learning models provide better forecast both on a static perspective (model train in 1973-2005 period and forecast on 2006 – 2018 period) and a dynamic perspective (initial model train in 1973-2005 period and rolling forecast with continuous re-training during the 2006 – 2018 period). Examining both error metrics and relevant plots it is evident that deep learning algorithms capture better the realized trends, especially in cases where the absence of linearities and the contemporaneous dependencies cause traditional modes to overshoot. 
	
	This first attempt at employing Deep Learning in macroeconomic time series forecasting shows that potential benefits may pave the ground for a wider spectrum of application in economic sciences. Of course, deep learning techniques even though they better address non-linear patterns they are not a panacea, especially in such  challenging problem as the prediction of the Sub-prime crisis in the US, but they certainly lie in the correct path. Criticism could rely on the black box nature of the algorithm which when compared to traditional econometrics does not provide a clear view of the economic relationships, but in any case, as it has been proven empirically,the economy is not dominated by clear linear patterns but from non-linear interactions which constantly evolve. Especially under the rolling forecast framework where both techniques follow the realized trend, one could use the results of the 2 techniques (Linear models – Deep Learning models) in a combined way, so that the linear model provides a first order approximation of the problem at hand revealing the economic rational and use also the more precise non-linear model to correct for temporal fluctuations.  

\newpage

\bibliographystyle{plain}
\bibliography{macro}

\clearpage
\newpage

\appendix
\section{Appendix}
\label{appendixa}
	
	\subsection{Static Forecast}

	\begin{figure}[!h]
	\centering
	\includegraphics[width=\linewidth]{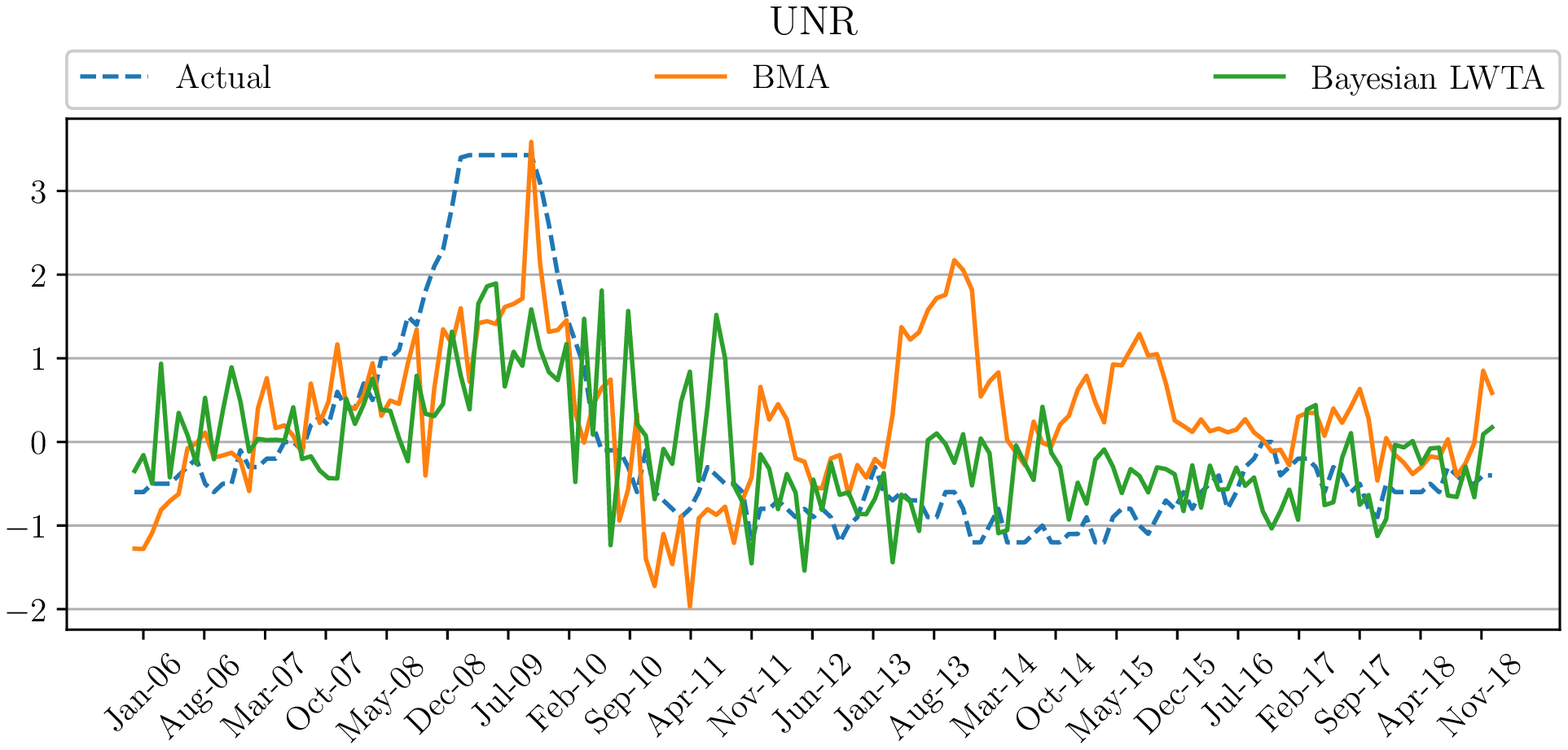}
	\caption{Unemployment Rate- Forecast Comparison. Actual value vs Bayesian Model Average (BMA) vs Bayesian Neural Network with Local Winner-Takes-All activation function (Bayesian LWTA).}
	\end{figure}

	\begin{figure}[!h]
		\centering
		\includegraphics[width=\linewidth]{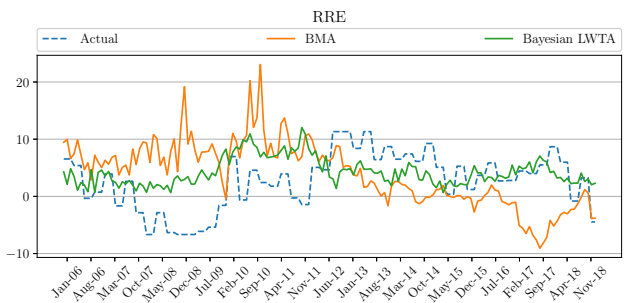}
		\caption{Real Estate price growth- Forecast Comparison. Actual value vs Bayesian Model Average (BMA) vs Bayesian Neural Network with Local Winner-Takes-All activation function (Bayesian LWTA).}
	\end{figure}

	\begin{figure}[!h]
		\centering
		\includegraphics[width=\linewidth]{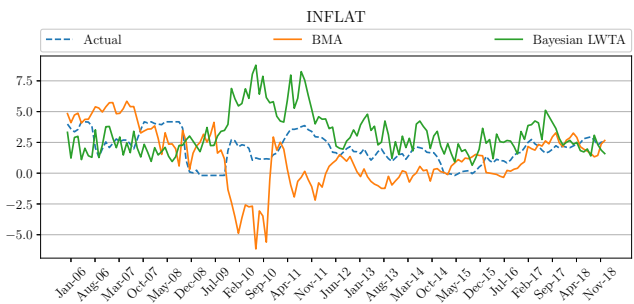}
		\caption{ Inflation- Forecast Comparison. Actual value vs Bayesian Model Average (BMA) vs Bayesian Neural Network with Local Winner-Takes-All activation function (Bayesian LWTA).}
	\end{figure}

	\begin{figure}[!h]
		\centering
		\includegraphics[width=\linewidth]{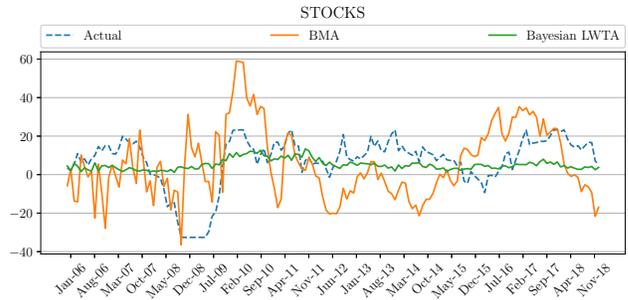}
		\caption{Stock market returns- Forecast Comparison. Actual value vs Bayesian Model Average (BMA) vs Bayesian Neural Network with Local Winner-Takes-All activation function (Bayesian LWTA).}
	\end{figure}

	\begin{figure}[!h]
		\centering
		\includegraphics[width=\linewidth]{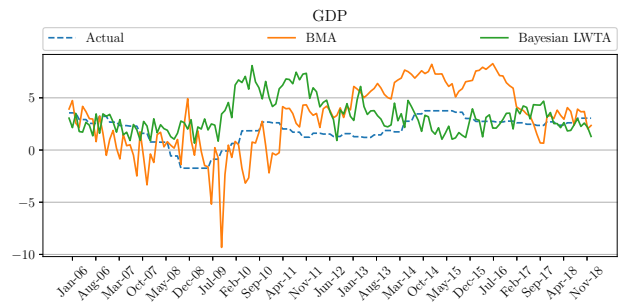}
		\caption{GDP growth- Forecast Comparison. Actual value vs Bayesian Model Average (BMA) vs Bayesian Neural Network with Local Winner-Takes-All activation function (Bayesian LWTA).}
	\end{figure}

	\clearpage
	\subsection{Rolling Forecast}

		\begin{figure}[!h]
		\centering
		\includegraphics[width=\linewidth]{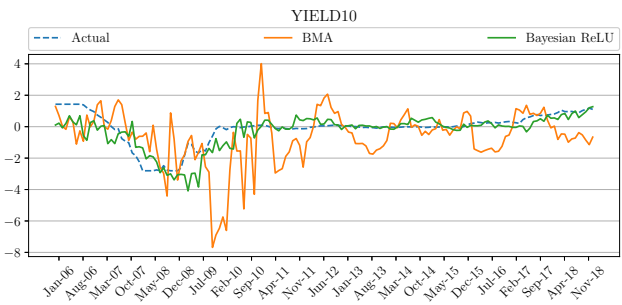}
		\caption{10 year Government bond yield change- Forecast Comparison. Actual value vs Bayesian Model Average (BMA) vs Bayesian Neural Network with ReLU activation function (Bayesian ReLU).}
		\label{fig:yeild10rolling}
	\end{figure}
	
	\begin{figure}[!h]
		\centering
		\includegraphics[width=\linewidth]{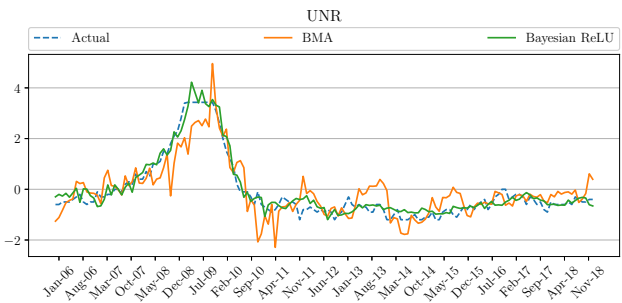}
		\caption{Unemployment Rate- Forecast Comparison. Actual value vs Bayesian Model Average (BMA) vs Bayesian Neural Network with ReLU activation function (Bayesian ReLU).}
	\end{figure}
	
	\begin{figure}[!h]
		\centering
		\includegraphics[width=\linewidth]{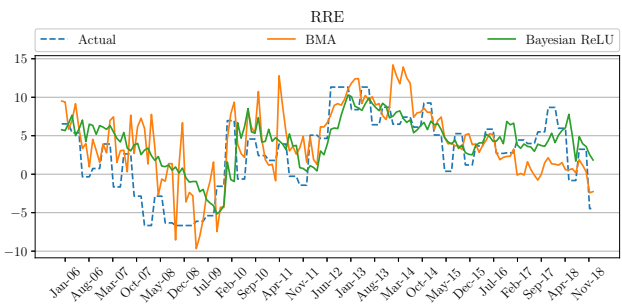}
		\caption{Real Estate price growth- Forecast Comparison. Actual value vs Bayesian Model Average (BMA) vs Bayesian Neural Network with ReLU activation function (Bayesian ReLU).}
	\end{figure}
	
	\begin{figure}[!h]
		\centering
		\includegraphics[width=\linewidth]{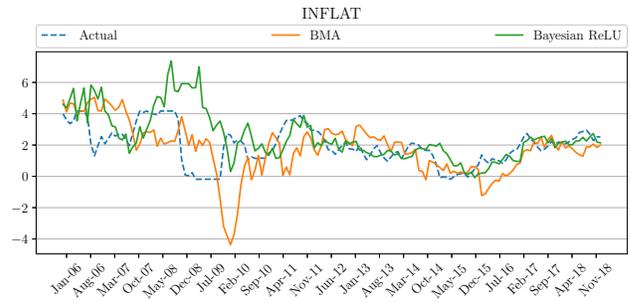}
		\caption{ Inflation- Forecast Comparison. Actual value vs Bayesian Model Average (BMA) vs Bayesian Neural Network with ReLU activation function (Bayesian ReLU).}
	\end{figure}
	
	\begin{figure}[!h]
		\centering
		\includegraphics[width=\linewidth]{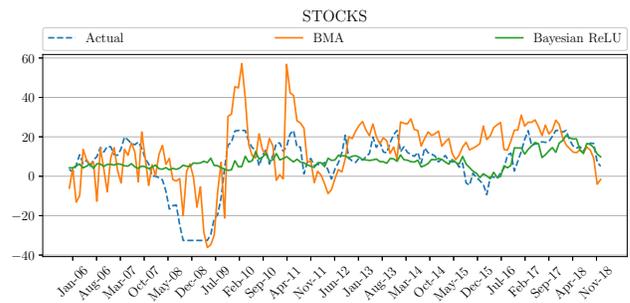}
		\caption{Stock market returns- Forecast Comparison. Actual value vs Bayesian Model Average (BMA) vs Bayesian Neural Network with ReLU activation function (Bayesian ReLU).}
	\end{figure}
	
	\begin{figure}[!h]
		\centering
		\includegraphics[width=\linewidth]{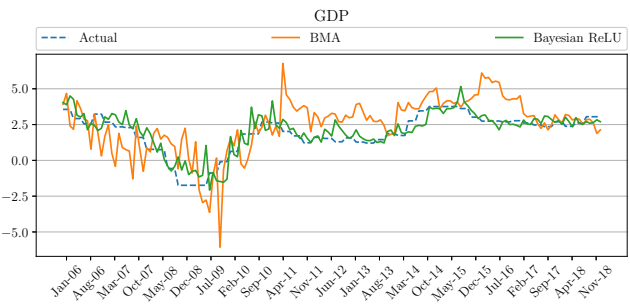}
		\caption{GDP growth- Forecast Comparison. Actual value vs Bayesian Model Average (BMA) vs Bayesian Neural Network with ReLU activation function (Bayesian ReLU).}
           \label{fig:GDPprolling}
	\end{figure}

\end{document}